# SUPPORTING LEVEL 1 PHYSICS & ASTRONOMY UNDERGRADUATES AT THE UNIVERSITY OF GLASGOW

Morag M. Casey[1]

[1]*Department of Physics & Astronomy, University of Glasgow, Scotland, UK.*

**Corresponding author's e-mail**: m.casey@physics.gla.ac.uk

## 1  Introduction

It is generally accepted that the retention and associated completion rates for first year classes are an area of concern for UK universities, and physics and astronomy classes at the University of Glasgow are exception. Classes are often large and, as result, student integration on academic and social levels can be difficult to achieve; some students perceive a lack of personal interest and support in what can be a stressful transition from secondary to tertiary education.

In order to address these issues, the author has been employed in a new departmental post, *Director of Learning Support for First Year*. The remit of this post is primarily the implementation of an improved personal contact and academic monitoring and support strategy for first year undergraduates.

The purpose of this paper is to present the ways in which it is hoped that the role of *Director of Learning Support* will positively impact on aspects of the forthcoming academic year.

## 2  Undergraduate Physics & Astronomy at the University of Glasgow

In order to understand how best to support Physics & Astronomy undergraduates, one must first understand the structure of science degrees at the University of Glasgow.

### 2.1 Undergraduate Science Degrees in Scotland

Under the Scottish schools' educational system [1], students can enter university after a minimum of twelve years of primary and secondary education. As a result Scottish undergraduate science degrees normally take a minimum of four years to complete. In comparison, students in English and Welsh universities are required to complete thirteen years of primary and secondary education before being admitted. In order to compete with universities in Europe, many UK universities now offer extended undergraduate degrees; these require students to undertake one further year of advanced study at undergraduate level.

One of the strengths of the longer time period required for undergraduate degrees at Scottish Universities is that students have the chance to study a broader range of subjects. In particular students may choose to study subjects completely unrelated to their degree major.

**2.2 Faculty Entry at the University of Glasgow**
In common with some other Scottish universities, the University of Glasgow grants students entry to a faculty rather than an individual course [2]; there are benefits and detriments to this system. As well as allowing students to study a broad-based curriculum, one benefit of the faculty entry system is that it allows students to change course at the end of their first year. The detriment to this system, from the perspective of individual departments, is that some students *do* decide to change course at the end of their first year.

The Department of Physics & Astronomy is part of the Faculty of Science in the University of Glasgow. Students granted entry to this faculty study three subjects in their first year; one common combination of subjects is physics, maths and chemistry. This combination allows students the option to pursue one of six different honours degree curricula. For example, physics can be studied on its own as a *single honours* degree or with one of the other subjects to complete a *combined honours* degree.

**2.3 Progression of Students from Level 1 to Level 2**
In order to be granted entry to the level 2 physics class, students must not only pass level 1 physics but also perform sufficiently well in their other two subjects. In particular, because level 2 physics students are required to complete level 2 maths, level 1 physics students must also perform sufficiently well in level 1 maths.

A first look at the data shows that in 2006-2007 (a typical year) 42.3% of the level 1 physics class enrolled in the level 2 physics class the subsequent year. However, this figure does not reveal the full complexity of the data and a more sophisticated analysis of the data is shown in Table 1.

**Table 1.** Cohort Analysis of 2006-2007 Physics 1Y Class

| Reason for Non-progression to Level 2 Physics | Percentage of L1 Class |
|---|---|
| Could not progress (did not study level 1 maths) | 13.3% |
| Did not intend to study physics beyond level 1 | 20.2% |
| Withdrew from level 1 physics during first year | 4.0% |
| Did not gain grades for entry to level 2 physics | 13.3% |
| Choose not to study level 2 physics | 7.0% |

The first two rows of data in *Table 1* provide supporting evidence of the effects of a faculty entry system. Large fractions of all level 1 science classes contain students who either cannot progress beyond level 1 in the subject because they do not take co-requisite subjects (such as level 1 maths for level 1 physicists) or never intend to study physics beyond level 1. Most students who withdrew from level 1 physics during their first year (third row) also withdrew from university entirely. It is the remaining two rows in *Table 1* that are, perhaps, of most interest.

*Students who did not gain grades for entry to level 2 physics*
In 50% of these cases, students did not gain the grades for entry to *any* level 2 subject. Students in this category either went on to re-enter level 1 anew or leave university completely. The other 50% did not gain grades for entry to level 2 physics but did gain grades for entry to level 2 in their other subjects. In a climate where financial constraints are high, it is perhaps not surprising that these students sought to cut their losses and pursue a degree curriculum in another subject.

***Students who chose to study something other than level 2 physics***
This category of students had stated an intention at the start of level 1 to pursue an honours curriculum in physics and, indeed, achieved the necessary grades to continue along that path. However, by the end of level 1, they had changed their minds and decided to pursue degrees in other subjects, evidence again of the effects of a faculty entry system.

It was with a desire to reduce the percentage of the class falling into these last two categories that the post of *Director of Learning Support* was created.

**3 The Role of Director of Learning Support for First Year**
The role of *Director of Learning Support* was created out of an attempt lower the percentage of the level 1 class that fall into the categories described in the last two rows of *Table 1*. The primary purpose of the role is to enhance the department's effective learning environments by identifying early-on "at-risk" students as well as providing extra targeted academic support.

**3.1  Defining the Role**
It was apparent from the start that the remit of the role ought to be well-defined; if not, there was the possibility of being swamped by requests for help. In terms of achieving the department's goal of producing good physics and astronomy honours graduates, it made sense to target primarily those students who professed a desire to continue beyond level 1. Looking at the assessment data from previous years it became apparent that it was with what might be termed the 'marginal-pass' students that any extra influence might be effective.

It was also agreed that the presentation of the role to both students and staff was going to be extremely important in determining its effectiveness. Therefore, it was determined that highlighting a specific subset of students based on what could be perceived as a negative criterion, "poor" assessment data, was to be avoided at all costs. This might run the risk of further isolating an already vulnerable group of students. In this light, it was determined that all first year physics and astronomy students should be able to make use of the extra support available. However, it is important to define measurable data from which any successes or failures of the role could be determined quantitatively.

**3.2  Early Warning System**
One aspect of the Scottish educational system is that many students are quite young when they enter university; most level 1 students in Scotland are 17-18 years old compared to those in England and Wales who are, on average, a year older. It has been recognised that many of the younger students find it difficult to cope with the transition from a well-structured secondary education to a world where they are expected to be entirely responsible adults. The assessment data from previous years provided evidence that some students failed level 1 physics not because they were intellectually incapable but because they did not submit required continuous assessment work on time. Discussions with students in higher years revealed that the failure to submit required work on time was usually due to the student being unaware that a deadline existed or that the work was required – a simple mismatch in communication between teaching staff and students.

Taking cues from work carried out at the University of Brighton [3] it was decided that the *Short Message Service (SMS)* of mobile telephones could be used to great effect. To avoid presenting the role of *Director of Learning Support* in a negative light, it was decided that students should not be required to provide a mobile telephone number, rather that they should be asked to *volunteer* it.

There are many benefits to using *SMS* to contact students, not least the ability to contact the student directly rather than leaving messages with flatmates or parents. Perhaps more subtle is the buffer that *SMS* allows the student compared to a telephone call. Departmental experience of direct telephone calls asking students for a required piece of work has rarely been described as a positive experience; on picking up the telephone most students engage in a knee-jerk reaction and claim that all is well. The buffer of the *SMS* allows students to compose a reasoned response and, more importantly, actually submit the work.

Therefore, it was decided that the *Director of Learning Support* would contact students by *SMS* immediately upon them missing any continuous assessment component of the class. This was completely different to the approach taken in previous years where students were contacted about missing work at the end of the first semester. For a student who had missed many deadlines, the end of the first semester could be too late for them to do anything constructive and, invariably, they failed the course. It was hoped, therefore, that contact early in the semester might avoid this situation and help more students pass.

**3.3 Extra Academic Support**
Traditional methods for extra academic support typically take the form of a series of extra revision classes or large-class tutorials. However, revision classes are notoriously badly attended and, even then, typically by the academically committed who will do well no matter what is thrown at them [4].

Taking cues from the work carried out at the University of St Andrews [5], it was decided that a more proactive approach was needed. All students would be able to make use an appointment-based one-to-one tuition system, run by a few members of staff. Pilot programs run in the honours astronomy class had been well-received by students. This approach appears to contradict the advice of [6] who suggests that effective learning support is best offered within the curriculum. However, by presenting the nature of this extra academic support as part of an integrated whole-class approach it is hoped that the role of *Director of Learning Support* will be perceived by students and staff in a wholly positive light.

**3.4 Future Work**
The approaches described in sections 3.2 and 3.3 will be rolled out for level 1 physics and astronomy classes in academic year 2007-2008. From this, it is hoped that a better understanding of level 1 data will be gleaned, allowing a streamlining of the role for academic year 2008-2009. The intention at that point is to expand the role of *Director of Learning Support* to include level 2 classes in physics and astronomy.

## 4  Conclusions

The interrogation of assessment data from level 1 undergraduate classes has revealed that the nature of students' continuation to honours degrees in physics and/or astronomy is complex. A large fraction of the level 1 classes do not continue to level 2 physics and/or astronomy classes either because they do not take the necessary co-requisite classes or have stated an intention to pursue other degree curricula from the start of level 1. The remaining subset of students who could proceed to level 2 but do not consists of two categories: students who do not gain the necessary grades to proceed and those who change direction and choose other subjects. The post of *Director of Learning Support* has been created to address the possibility of lowering the percentage of the level 1 classes in these two categories.

It is hoped that the role of *Director of Learning Support* will impact positively in two main areas. The early identification of students who are struggling to keep up with continuous assessment work should allow positive action to be taken in order to help them complete the class successfully. A system of extra targeted academic support will also be implemented to help marginal-pass students improve their grades and, hopefully, their overall experience of level 1 classes in the Department. It is intended that this role be extended to include level 2 classes in future.